# Carrier Dynamics and Transient Photobleaching in Thin Layers of Black Phosphorus

Ryan J. Suess,[1,2,a)] Mohammad M. Jadidi,[1,2] Thomas E. Murphy,[1,2] and Martin Mittendorff[1]

[1]*Institute for Research in Electronics & Applied Physics, University of Maryland, College Park, Maryland, 20742, USA*

[2]*Department of Electrical & Computer Engineering, University of Maryland, College Park, Maryland, 20742, USA*

We present polarization-resolved transient transmission measurements on multi-layer black phosphorus. Background free two-color pump-probe spectroscopy measurements are carried out on mechanically exfoliated black phosphorus flakes that have been transferred to a large-bandgap, silicon carbide substrate. The blue-shifted pump pulse (780 nm) induces an increased transmission of the probe pulse (1560 nm) over a time scale commensurate with the measurement resolution (hundreds of fs). After the initial pump-induced transparency, the sign of the transient flips and a slower enhanced absorption is observed. This extended absorption is characterized by two relaxation time scales of 180 ps and 1.3 ns. The saturation peak is attributed to Pauli blocking while the extended absorption is ascribed to a Drude response of the pump-induced carriers. The anisotropic carrier mobility in the black phosphorus leads to different weights of the Drude absorption, depending on the probe polarization, which is readily observed in the amplitude of the pump-probe signals.

a) ryan.suess@gmail.com

Black phosphorus is a relatively new member in the family of two-dimensional materials that can be mechanically exfoliated to atomically thin films. Its crystalline structure forms a corrugated hexagonal lattice that exhibits unique anisotropic optical and electrical properties [1] that have recently been experimentally observed [2]. The direct bandgap in black phosphorus is dependent on the number of layers, ranging from about 0.3 eV in bulk (5 or more layers) [3] to about 2 eV for a single layer [4]. In contrast to other two-dimensional materials, where the mobility is commonly far below that of graphene [5,6,7], the mobility in black phosphorus can approach values on the order of 10,000 $cm^2V^{-1}s^{-1}$ [1]. The presence of a direct band gap along with the high carrier mobility makes black phosphorus an interesting candidate for fast field-effect transistors [8] and optoelectronic devices like photodetectors [9,10]. Similar to graphene, black phosphorus also features a strong saturable absorption [11], a feature that has already been exploited for use in mode-locked lasers [12]. Polarization-resolved linear spectroscopy measurements have shown that the optical absorption is strongly anisotropic [ 13 ], while photoluminescence measurements reveal an even stronger anisotropy, with the photoemission being strongly polarized (97%) in the direction perpendicular to the corrugations in the material [14]. Apart from linear absorption and electrical transport studies, the carrier dynamics in black phosphorus have not been thoroughly investigated. Two very recent studies report pump-probe measurements on black phosphorus flakes on a silicon substrate using a reflection geometry [15,16]. In this paper we present a polarization-resolved, collinear near-infrared pump-probe study on black phosphorus flakes in transmission. To achieve a background free transmission measurement, a semi-insulating, large bandgap silicon carbide substrate is used.

The two-color pump-probe experiments were carried out on multi-layer black phosphorus flakes obtained by mechanical exfoliation and transferral onto a 300 μm thick silicon carbide substrate. This substrate was chosen since it possesses a bandgap nearly two (four) times that of the pump (probe) photon energy; thus allowing for a pump-probe signal from the black phosphorus flake that is unaffected by contributions from the substrate. This property of the substrate was confirmed by pump-probe measurements on the bare silicon carbide where no signals could be observed. The exfoliation process yielded flakes between 30-80 layers and several hundreds of μm$^2$ in area, as determined by optical contrast microscopy [17]. The results reported here are for a flake with 10 % linear transmission and an estimated thickness of approximately 80 layers. As a means of carrying out precision photometry, we patterned a circular chromium aperture that covers all but a 10 μm diameter region in the center of the black phosphorus flake under investigation. This ensures uniform illumination of the irregular black phosphorus flakes and guarantees that only probe light transmitted through the black phosphorus is measured by the detector. The aperture was patterned using electron-beam lithography and liftoff. Following fabrication, a 200 nm PMMA cover layer was spin-coated over the entire device to prevent degradation of the black phosphorus which is known to react with ambient humidity [17].

The measurements were carried out at room temperature (300 K) using two fiber lasers (Menlo Systems). The lasers produce approximately 100 fs wide pulses at a repetition rate of 100 MHz and center wavelengths of 780 nm and 1560 nm for the pump and probe, respectively. The two lasers were electronically synchronized with a small difference in repetition rate, allowing for large pump-probe delays (up to 10 ns) to be achieved without the use of a mechanical delay line. Both the pump and probe separately pass through a half-wave plate to allow the polarization of each beam to be independently controlled. Figure 1(a) defines the polarization (shown as double-ended arrows) with respect to the major and minor axis of the measured relative linear transmission ellipse at 1560 nm for the particular black phosphorus flake presented in this work. The difference in transmission is caused by the anisotropy of the band structure [18,19] and provides a means to establish the flake orientation prior to carrying out the polarization-resolved pump-probe measurements. In Fig. 1(a), the major axis (along the 0° direction) corresponds to horizontal polarization while the minor axis (along the 90° direction) corresponds to vertical polarization. Diagonal polarization refers to a linear polarization angle of 45° and 135°. The inset in Fig. 1(a) shows the linear polarization angle with respect to the corrugation orientation in black phosphorus. The main experimental diagram in Fig. 1(b) defines the independent linear polarization angles for both the pump and probe pulses as $\Phi_P$ and $\Phi_p$, respectively. The pump is chopped (550 Hz) and combined with the probe beam before being reflected into an aspheric objective (NA = 0.4) using a beam splitter. The light passing through the lens focusses the beams onto the sample, which is positioned such that the pump and probe beams overfill the aperture. The light transmitted through the aperture is recollimated by a second objective, passed through an intrinsic silicon window to filter out the pump, and directed into the signal input of a balanced detector unit. The remaining light transmitted through the beam splitter passes through a variable optical attenuator, intrinsic silicon window, and into the reference input of the balanced detector. A lock-in amplifier monitors the pump-induced modulation of the probe in the black phosphorus flake as the delay between the pump and probe is scanned. For all measurements the probe fluence was approximately three orders of magnitude smaller than that of the pump. Figure 1(b) shows a diagram of the sample illumination geometry, an energy band diagram for the experiment, and an optical micrograph of a black phosphorus-filled aperture.

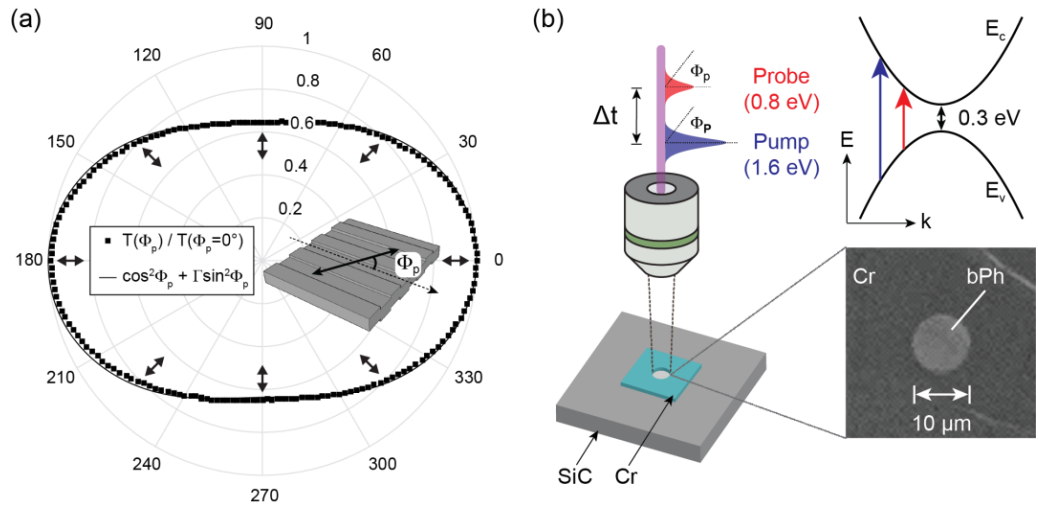

Figure 1: (a) Relative change in linear transmission as a function of the linear polarization angle with phenomenological fit. The double-ended arrows show the direction of the electric field oscillation and define horizontal, diagonal, and vertical polarizations. The inset shows the polarization angle with respect to the corrugated surface of the black phosphorus. (b) In clockwise order: Diagram showing the sample illumination geometry and definition of the pump-probe delay, simplified band diagram of bulk black phosphorus showing the relevant energy scales in the experiment, and an optical micrograph of a black phosphorus(bPh)-filled aperture.

To investigate the power dependence of the pump induced changes, transient transmission data were collected for pump fluences ranging from 3 µJ/cm$^2$ to 80 µJ/cm$^2$ for vertically polarized pump and probe. The first 60 ps of the pump-probe signals are shown in Fig. 2(a). The pump-probe traces are characterized by an initial pump-induced transparency having a duration on the same time scale as the pump and probe pulse overlap, followed by a slower transient absorption. In the inset of Fig. 2 (a) the $\Delta T/T_0$ values from the main panel of Fig. 2(a) are plotted for fixed delay values $\Delta t = 0$ ps and 10 ps as a function of pump fluence. The transmission spectra for these two phenomenologically distinct regions of the transient response (i.e., the absorption saturation region near $\Delta t = 0$ ps, and the reduced transmission region for $\Delta t > 5$ ps) scale linearly over nearly two orders of magnitude in pump fluence. To evaluate the relaxation dynamics, Fig. 2(b) plots the carrier-induced transient absorption on a logarithmic scale over a 5 nanosecond timescale, along with a bi-exponential fit to the data. For the lowest pump fluence (grey data set in Fig. 2), the relaxation can be described by a single exponential decay with a time constant of 770 ps. With increasing pump-power, a faster time constant of 180 ps (± 30 ps) becomes more prominent (visible as an increasingly large upward curvature of the data at smaller delays as the fluence is increased), while the slower time constant linearly increases to 1.8 ns at 80 μJ/cm$^2$ (appearing as a decrease in the slope of the data as the fluence is increased). We attribute the transient pump-induced transparency to Pauli blocking, while the subsequent slower negative transient reflects the lifetime of the pump-generated free carrier population. Absorbed pump photons create free-carriers in the black phosphorus that will increase the conductivity and lead to both an increased absorption and reflection [13]. We note that both effects could contribute to the observed negative transient seen here.

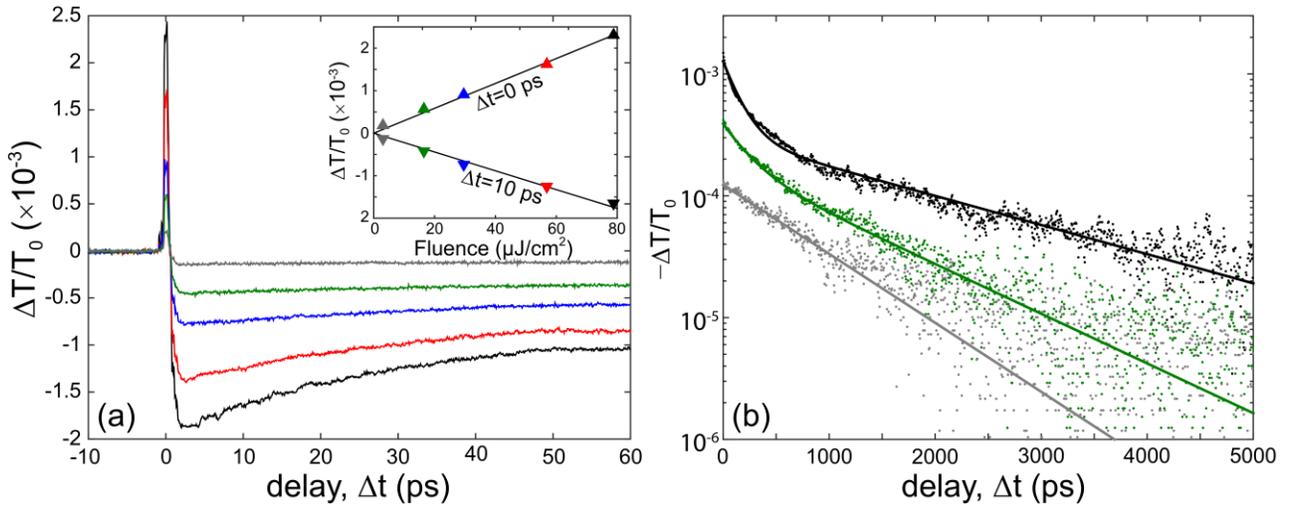

Figure 2: (a) Change in transmission as a function of probe delay for a variety of pump powers. The inset shows the linearity of the change in transmission with pump fluence at 0 ps and 10 ps. (b) Logarithmic plot of $-\Delta T/T_0$ for three pump fluences along with bi-exponential fits (solid lines).

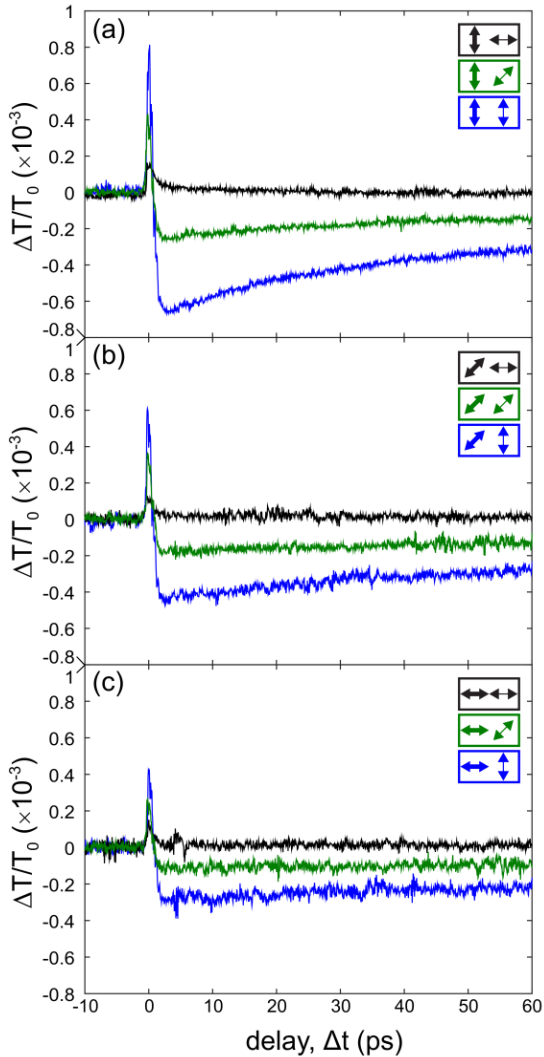

Figure 3: Polarization-resolved pump-probe signals. Each panel shows a set of curves for a fixed pump and varied probe polarization. In the legend, the thick (thin) arrow designates the pump (probe) polarization.

To further investigate the carrier dynamics, we performed a set of pump-probe measurements for different combinations of pump and probe polarization. The pump and probe polarizations were set to the four polarization angles of 0°, 45°, 90°, and 135° (corresponding to horizontal, diagonal, vertical, and diagonal polarization). The measurements for pump and probe polarization angles of 135° were performed as a crosscheck and yielded identical results to the measurements carried out at 45°. Figure 3 presents the pump-probe transmission spectra for 9 unique polarization configurations. Each panel of Fig. 3 contains the traces taken at different probe polarizations for a fixed pump polarization. All measurements were taken at a constant incident pump fluence of 30 µJ/cm². By comparing Fig. 3(a) and Fig. 3(c), one sees that in all cases the transient response is smaller when the pump is vertically polarized than when it is horizontally polarized. This dependence on pump polarization cannot be fully explained by polarization-dependent absorption of the pump, estimated to be only about 5% for the multilayer black phosphorus considered here. Assuming that the positive peak is caused primarily by Pauli blocking, this suggests that the carriers are changing the orientation of their momentum, aligning along the energy gradient of the anisotropic band structure. The short time scale on which this transient enhancement of transmission occurs (within the resolution of the measurement) implies an efficient carrier-phonon scattering by which the carriers minimize their energy within the conduction and valence band, respectively.

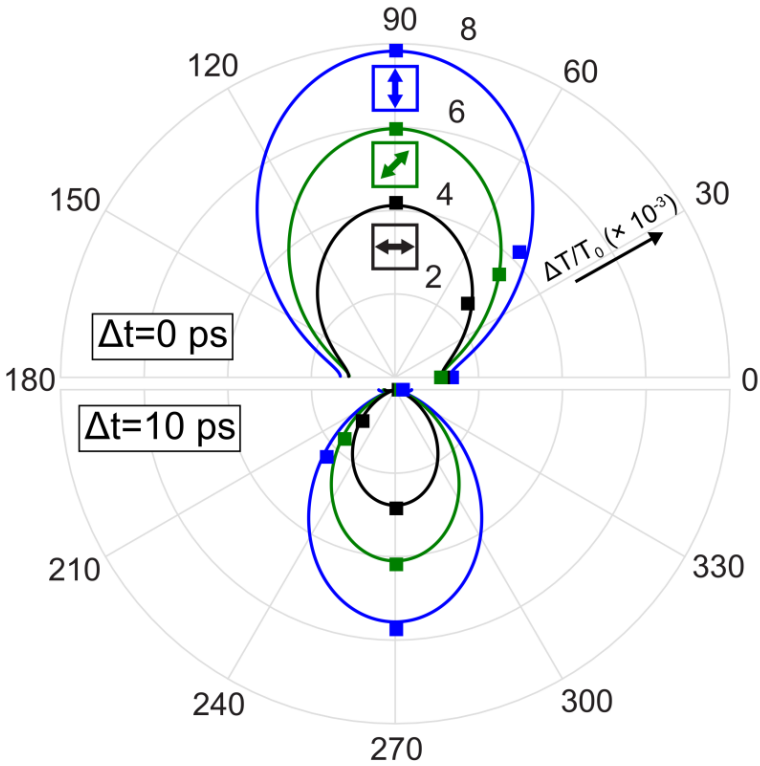

Figure 4: $\Delta T/T_0$ at 0 ps (upper half-circle) and 10 ps (lower half-circle) for all pump-probe polarization configurations as a function of probe polarization angle. The colored arrows indicate the pump polarization. The solid lines are phenomenological fits of the form: $a\cos^2\Phi_p + b\sin^2\Phi_p$.

We attribute the extended negative transient to intraband (Drude) absorption, resulting from an excess pump-induced carrier population, similar to what has been reported recently in graphene [20]. For a better understanding of the anisotropy of the signal, the amplitude of the positive peak at 0 ps and the amplitude of the negative transmission at a time delay of 10 ps is plotted in Fig. 4. While both components of the transient response depend on the probe polarization, the slower negative transient response shows a much more pronounced anisotropy, and is nearly absent when the probe is horizontally polarized. The stronger anisotropy of the delayed absorption is likely explained by the large differences in the carrier mobility in black phosphorus. The mobility in few-layer black phosphorus has been recently calculated to be around 400 and 1,500 $cm^2V^{-1}s^{-1}$ for electrons moving parallel or perpendicular to the corrugations, respectively [1], which could explain the observed anisotropic transient response. Additional theoretical modelling is needed to fully understand the mechanisms governing the polarization anisotropy of the pump-induced change in transmission.

In summary, two-color pump-probe experiments were conducted on thin layers of black phosphorus in a transmission geometry. We observed a transient photobleaching on the time scale of the temporal resolution of the measurement followed by a long-lived pump-induced absorption exhibiting two characteristic time scales on the order of 180 ps and 1.3 ns. Polarization-resolved measurements revealed an anisotropy that was stronger for the time delayed portion of the response and is attributed to the strong anisotropy of the electron mobility in black phosphorus. These results emphasize the importance of polarization and black phosphorus orientation in optoelectronic devices as well as purely optical devices that make use of the material's nonlinear optical properties.

Parts of this work were supported by the Office of Naval Research (ONR) award number N000141310865 and the National Science Foundation (NSF) award number ECCS1309750. The sample fabrication was carried out at the University of Maryland Nanocenter.